\documentclass[useAMS,usenatbib,usegraphicx]{mn2e}
\usepackage{graphicx, epstopdf, txfonts, color}

\defcitealias{paper1}{VJH10}

\title[Exoplanet Magnetic Fields and Bow Shocks]{Prospects for Detection of Exoplanet Magnetic Fields Through Bow-Shock Observations During Transits}
\author[Vidotto, Jardine, \& Helling]{A. A. Vidotto \thanks{E-mail: Aline.Vidotto@st-andrews.ac.uk}, M. Jardine, Ch. Helling \\
SUPA, School of Physics \& Astronomy, University of St Andrews, North Haugh, St Andrews,  KY16 9SS, UK\\
}
\begin{document}
\date{Accepted ??. Received ??; in original form ??}
\pagerange{\pageref{firstpage}--7} \pubyear{2010}
\maketitle
\label{firstpage}

\begin{abstract}
An asymmetry between the ingress and egress times was observed in the near-UV light curve of the transit planet WASP-12b. Such asymmetry led us to suggest that the early ingress in the UV light curve of WASP-12b, compared to the optical observations, is caused by a shock around the planet, and that shocks should be a common feature in transiting systems. Here, we classify all the transiting systems known to date according to their potential for producing shocks that could cause observable light curve asymmetries. We found that $36/92$ of known transiting systems would lie above a reasonable detection threshold and that the most promising candidates to present shocks are: WASP-19b, WASP-4b, WASP-18b, CoRoT-7b,  HAT-P-7b, CoRoT-1b, TrES-3, and WASP-5b. For prograde planets orbiting outside the co-rotation radius of fast rotating stars, the shock position, instead of being ahead of the planetary motion as in WASP-12b, trails the planet. In this case, we predict that the light curve of the planet should present a late-egress asymmetry. We show that CoRoT-11b is a potential candidate to host such a behind shock and show a late egress. If observed, these asymmetries can provide constraints on planetary magnetic fields. For instance, for a planet that has a magnetic field intensity similar to Jupiter's field ($\sim 14$~G) orbiting a star whose magnetic field is between $1$ and $100$~G, the stand-off distance between the shock and the planet, which we take to be the size of the planet's magnetosphere, ranges from $1$ to $40$ planetary radii.  
\end{abstract}

\section{Introduction}
Recently, \citet{fossati2010} observed that the near-UV transit light curve of the close-in giant planet WASP-12b shows an early ingress as compared to its optical transit. Such observations were interpreted as due to the presence of asymmetries in the exosphere of the planet \citep[e.g.,][]{lai2010}. In particular, we showed that this asymmetry could be explained by the presence of a shock around the planet \citep*[][hereafter, VJH10]{paper1}. 

By analysing which characteristics of the ambient surrounding the planet could lead to an observable early UV ingress, we showed that shock formation is likely to be a common feature of transiting systems. Furthermore, once the stand-off distance of the shock (determined through the time difference of transit observations at different wavelengths) and the stellar magnetic field strength are known, we can derive the planetary magnetic field intensity. The planetary magnetic field is believed to be responsible for shielding the planet against the erosion of the planetary atmosphere by the host star's wind or the impact of energetic cosmic particles. Such effects could harm creation and development of life in the planet. Furthermore, the presence of a planetary magnetic field may induce star-planet interactions, e.g., through reconnection between stellar and planetary magnetic field lines \citep{stevens2005}. 

In this Letter, we extend the idea developed for WASP-12b \citepalias{paper1} to all known transiting exoplanets and investigate cases in which we expect to see light curve asymmetries. We discuss a selection of transiting planets for which we expect a shock to form (Section~\ref{sec.shock}). This shock can either cause an observable early ingress or late egress during transition (Section~\ref{sec.geometry}), depending on the characteristics of the ambient surrounding the planet, as well as on the planetary orbital properties. Conclusions are presented in Section~\ref{sec.conclusion}.

\section{Which planets might develop shocks?}\label{sec.shock}
As in \citetalias{paper1}, we discuss two possible scenarios for shock formation: (1) the stellar magnetic field is strong enough to confine the coronal plasma to co-rotate out to the orbital radius of the planet; (2) the stellar magnetic field is not sufficiently strong, and the coronal material escapes in the form of a stellar wind. The data used here were a selection of transiting systems from the exoplanets encyclopedia (http://exoplanet.eu/catalog.php), except for the sky-projected stellar rotation velocity $v \sin(i)$, which was taken from \citet{2010ApJ...719..602S}. In this Letter, we disregard planets that have either high eccentricity ($>0.3$) or far orbits ($> 0.1$~AU).

\subsection{Co-rotating hydrostatic corona}
For a shock to develop, the relative velocity between the planet and the surrounding stellar corona must be greater than the local sound speed \citepalias{paper1}: $|u_{\varphi, {\rm cor}} - u_K| > c_s$. Here, $u_{\varphi, {\rm cor}} = [v \sin(i)/\sin(i)] (R_{\rm orb}/R_*) $ is the azimuthal velocity of the co-rotating corona at the orbital radius of the planet $R_{\rm orb}$, $c_s= (k_B T/m)^{1/2}$ the sound speed, $m=\mu m_p$ the mean particle mass (we adopt $\mu=0.66$), $m_p$ the proton mass, $k_B$ the Boltzmann constant, and $u_K=(G M_*/R_{\rm orb})^{1/2}$ is the Keplerian velocity of the planet that orbits a star with mass $M_*$ and radius $R_*$. Since the sound speed depends on the coronal temperature $T$, we can write this as the maximum temperature for shock formation
\begin{equation}\label{eq.tmax}
T_{\rm max} = {( u_{\varphi, {\rm cor}} - u_K)^2 m}/{k_B}.
\end{equation}
If the temperature of the local medium exceeds this value, then no shock will form. Table~\ref{table1} shows the maximum temperature for all the planets in our sample (an extended version of Table~\ref{table1} is available online). We note that, for stars where we do not know the stellar rotation rate, we take the limiting case of zero stellar rotation and so $u_{\varphi, {\rm cor}} \to 0$, resulting  in $T_{\rm max} \to T_{\rm crit}$, where $T_{\rm crit}$ is given by
\begin{equation}\label{eq.tcrit}
T_{\rm crit} = \frac{G M_* m}{R_{\rm orb} k_B} = 23.1 \times 10^6 \mu \frac{M_* / M_\odot}{R_{\rm orb} / R_\odot} {\rm K}.
\end{equation}
We note that the maximum temperature for shock formation for the majority of the planets is $(1$ -- $4) \times 10^{6}~$K. Although stellar flares can raise the coronal plasma temperature above these values, it is unlikely that a corona would be hotter than this.

\begin{table*}
\caption{Transiting planetary systems known as of September/2010. Planets are ordered in increasing coronal densities of their host stars, taken as a proxy for the detection of light curves asymmetries. The columns are: (1) the planet name, (2) mass, (3) radius, (4) orbital period, and (5) semi-major axis, (6) the distance to the system, (7) the host star spectral type, (8) mass, and (9) radius, (10) the sky-projected stellar rotation velocity, (11) the maximum temperature required for shock formation; (12) the local density around the planet for the confined corona, and (13) considering the coronal density scales with $\Omega_*$, (14) the size of the planet magnetosphere for a planet with $B_p=14$~G and a star with $B_*=1$~G, (15) the same but for $B_*=100$~G, (16) the minimum planetary magnetic field relative to the stellar one ($f = (B_p/B_*)_{\rm min}$) that is required to sustain a magnetosphere. An extended version of this table is available online.} \label{table1}
\begin{center}
\begin{tabular}{lccccccccccccccc}
\hline
Planet&$M_p$&$R_p$&$P_{\rm orb}$&$R_{\rm orb}$&$d$&Spec.&$M_*$&$R_*$&$v \sin(i)$&$T_{\rm max}$&$\log\left[\frac{n}{{\rm cm}^{-3}}\right]$&$\log\left [\frac{n}{{\rm cm}^{-3}}\right]$&	$r_M/R_p$&$r_M/R_p$&$f$	\\
Name&$(M_{\rm J})$&$(R_{\rm J})$&(d)&(AU)&(pc)&Type&$(M_\odot)$&$(R_\odot)$&(km/s)&(MK)&unsc.&scaled&($1$G)&($100$G)&$(\%)$\\
\scriptsize{(1)} & \scriptsize{(2)} & \scriptsize{(3)} & \scriptsize{(4)} & \scriptsize{(5)} & \scriptsize{(6)} & \scriptsize{(7)} & \scriptsize{(8)}& \scriptsize{(9)} & \scriptsize{(10)}  & \scriptsize{(11)}  & \scriptsize{(12)} & \scriptsize{(13)} & \scriptsize{(14)}& \scriptsize{(15)}& \scriptsize{(16)}\\
\hline \hline
WASP-12b	&	$	1.41	$	&	$	1.79	$	&	$	1.09	$	&	$	0.023	$	&	$	267	$	&	G0	&	$	1.35	$	&	$	1.57	$	&	$	2.2	$	&	$	3.96	$	&	$	7.02	$	&	$	6.86	$	&	$	7.5	$	&	$	1.6	$	&	$	3.2	$	\\
OGLE-TR-56b	&	$	1.30	$	&	$	1.20	$	&	$	1.21	$	&	$	0.023	$	&	$	1500	$	&	G	&	$	1.17	$	&	$	1.32	$	&	$	3.2	$	&	$	3.32	$	&	$	6.71	$	&	$	6.80	$	&	$	8.8	$	&	$	1.9	$	&	$	2.0	$	\\
WASP-19b	&	$	1.15	$	&	$	1.31	$	&	$	0.79	$	&	$	0.016	$	&	$	-	$	&	G8V	&	$	0.95	$	&	$	0.93	$	&	$	4	$	&	$	3.61	$	&	$	6.62	$	&	$	6.95	$	&	$	9.1	$	&	$	2.0	$	&	$	1.8	$	\\
SWEEPS-11 	&	$	9.70	$	&	$	1.13	$	&	$	1.80	$	&	$	0.030	$	&	$	8500	$	&	$-$	&	$	1.10	$	&	$	1.45	$	&	$	-	$	&	$	2.62	$	&	$	6.51	$	&	$	-	$	&	$	10.7	$	&	$	2.3	$	&	$	1.1	$	\\
WASP-4b	&	$	1.12	$	&	$	1.42	$	&	$	1.34	$	&	$	0.023	$	&	$	300	$	&	G7V	&	$	0.90	$	&	$	1.15	$	&	$	2	$	&	$	2.55	$	&	$	6.43	$	&	$	6.37	$	&	$	10.3	$	&	$	2.2	$	&	$	1.3	$	\\
WASP-18b	&	$	10.43	$	&	$	1.17	$	&	$	0.94	$	&	$	0.020	$	&	$	100	$	&	F9	&	$	1.28	$	&	$	1.23	$	&	$	11	$	&	$	3.11	$	&	$	6.40	$	&	$	7.05	$	&	$	8.6	$	&	$	1.9	$	&	$	2.2	$	\\
CoRoT-7b	&	$	0.02	$	&	$	0.15	$	&	$	0.85	$	&	$	0.017	$	&	$	150	$	&	K0V	&	$	0.93	$	&	$	0.87	$	&	$	3.5	$	&	$	3.36	$	&	$	6.38	$	&	$	6.69	$	&	$	10.2	$	&	$	2.2	$	&	$	1.3	$	\\
CoRoT-14b	&	$	7.60	$	&	$	1.09	$	&	$	1.51	$	&	$	0.027	$	&	$	1340	$	&	F9V	&	$	1.13	$	&	$	1.21	$	&	$	-	$	&	$	2.99	$	&	$	6.36	$	&	$	-	$	&	$	11.5	$	&	$	2.5	$	&	$	0.91	$	\\
HAT-P-7b	&	$	1.80	$	&	$	1.42	$	&	$	2.20	$	&	$	0.038	$	&	$	320	$	&	$-$	&	$	1.47	$	&	$	1.84	$	&	$	3.8	$	&	$	2.29	$	&	$	6.21	$	&	$	6.22	$	&	$	10.6	$	&	$	2.3	$	&	$	1.2	$	\\
OGLE-TR-132b	&	$	1.17	$	&	$	1.25	$	&	$	1.69	$	&	$	0.031	$	&	$	1500	$	&	F	&	$	1.26	$	&	$	1.34	$	&	$	5	$	&	$	2.24	$	&	$	5.78	$	&	$	6.05	$	&	$	11.8	$	&	$	2.5	$	&	$	0.84	$	\\
CoRoT-1b	&	$	1.03	$	&	$	1.49	$	&	$	1.51	$	&	$	0.025	$	&	$	460	$	&	G0V	&	$	0.95	$	&	$	1.11	$	&	$	5.2	$	&	$	1.98	$	&	$	5.71	$	&	$	6.08	$	&	$	11.8	$	&	$	2.6	$	&	$	0.84	$	\\
TrES-3 	&	$	1.91	$	&	$	1.31	$	&	$	1.31	$	&	$	0.023	$	&	$	-	$	&	G	&	$	0.92	$	&	$	0.81	$	&	$	1.5	$	&	$	2.65	$	&	$	5.63	$	&	$	5.60	$	&	$	14.3	$	&	$	3.1	$	&	$	0.47	$	\\
WASP-5b	&	$	1.64	$	&	$	1.17	$	&	$	1.63	$	&	$	0.027	$	&	$	297	$	&	G4V	&	$	1.02	$	&	$	1.08	$	&	$	3.5	$	&	$	2.15	$	&	$	5.63	$	&	$	5.84	$	&	$	13.0	$	&	$	2.8	$	&	$	0.63	$	\\
\hline
\end{tabular}
\end{center}
\end{table*}

\begin{figure}
	\includegraphics[width=84mm]{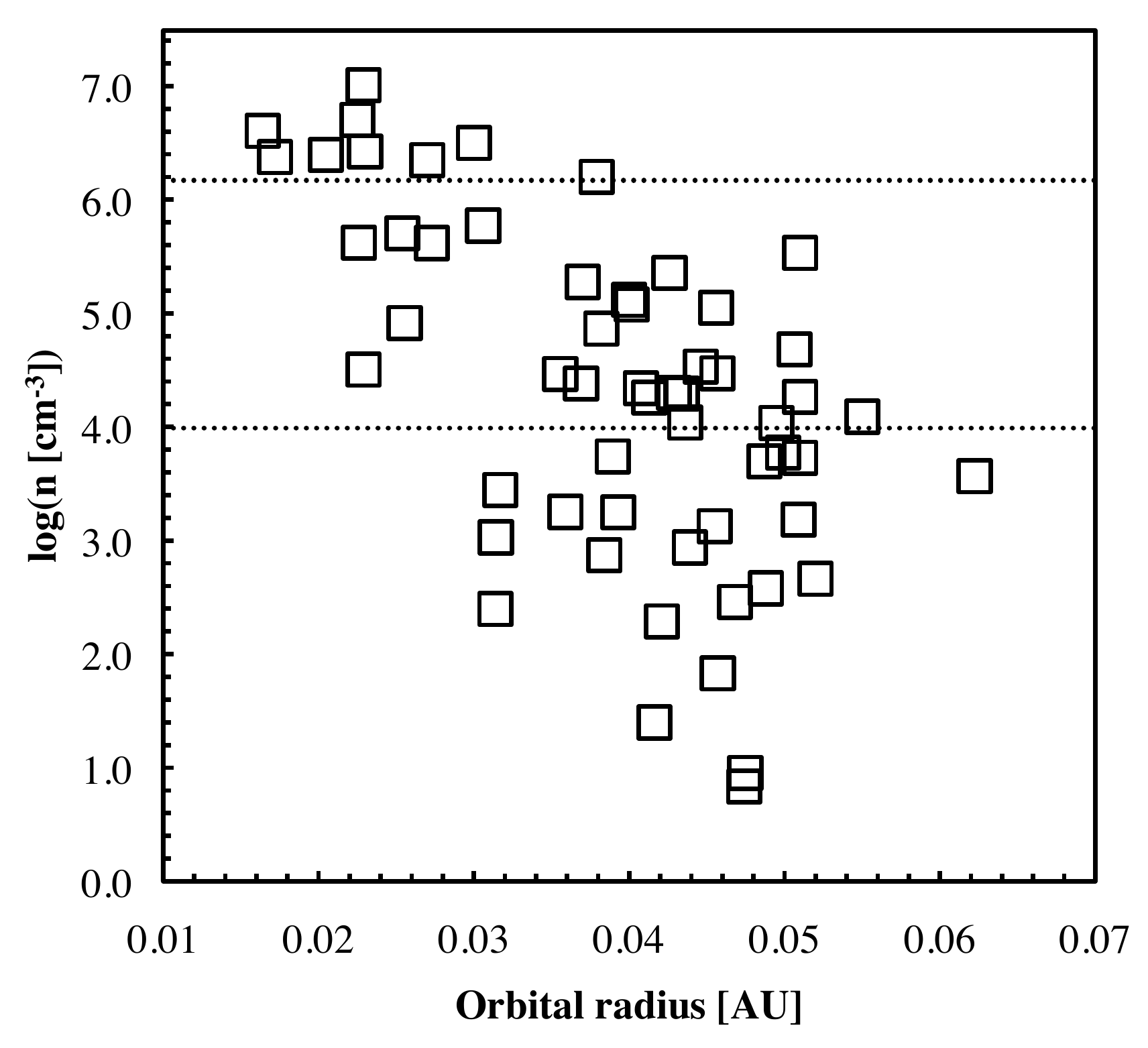}
\caption{Predicted coronal density at the orbit of each planet, assuming that the stellar magnetic field is strong enough to confine the coronal gas out to the orbit of the planet. 
We assume all the stars have the same solar coronal base density. The upper dashed line represents the minimum density value derived for WASP-12b \citepalias[$n_{\rm obs} = 1.5\times 10^6$~cm$^{-3}$, ][]{paper1}. The lower dashed line represents a lower limit for the density given the uncertainties in abundances, cross-sections and geometric depths through the shock. Approximately, $40\%$ of the planets in our sample lie above the detection limit.}\label{dens_all}
\end{figure}

For a shock to be detected, it must compress the local plasma to a density high enough to cause an observable level of optical depth. For a hydrostatic, isothermal corona, the local density is
\begin{equation}\label{eq.dens-hyd}
 n(R_{\rm orb}) = n_0\exp\left[ \frac{GM_*/R_*}{k_B T/m} \left( \frac{R_*}{R_{\rm orb}} -1 \right) + \frac{[v \sin(i)]^2/2}{k_B T/m} \left( \frac{R_{\rm orb}^2}{R_*^2}-1 \right)\right] ,
\end{equation}
where $n_0$ is the density at the base of the corona. If we now consider $T=T_{\rm max}$, we can obtain a minimum density (column 12 of Table~\ref{table1}). Figure~\ref{dens_all} shows this critical density as a function of $R_{\rm orb}$ for a range of planets, assuming that all stars have a base coronal density equal to that of the Sun \citep[$n_{0,\odot} = 10^8~{\rm cm}^{-3}$,][]{withbroe1988}. For comparison, we show in Figure~\ref{dens_all} the minimum density ($n_{\rm obs} = 1.5\times 10^6$~cm$^{-3}$) derived for WASP-12b \citepalias[upper dashed line, ][]{paper1}. The main uncertainties in the determination of the local density surrounding the planet come from the plasma abundances, cross-sections and geometric depths through the shock. The geometric effects alone allow for one order of magnitude uncertainty. We allow one further order of magnitude for the other effects. To conservatively account for these uncertainties in the density, we present in Figure~\ref{dens_all} a lower limit for the density (lower dashed line) that could still provide detection ($n_{\rm min} \simeq 10^4$~cm$^{-3}$). We note that $36$ out of $92$ planets lie above a reasonable detection threshold, suggesting that a detectable shock might indeed be a common feature surrounding transiting planets.

Some planets orbit extremely close to the co-rotation radius of the star. At this radius, $u_K \approx u_{\varphi, {\rm cor}}$ and $\Delta u \approx 0$. Therefore, in these cases, namely, XO-3b, CoRoT-3b, CoRoT-4b, and CoRoT-6b, no shock will be developed. However, if these planets are embedded in the wind of their host stars, which does not co-rotate with the star, a shock may be formed.

\begin{figure}
	\includegraphics[width=84mm]{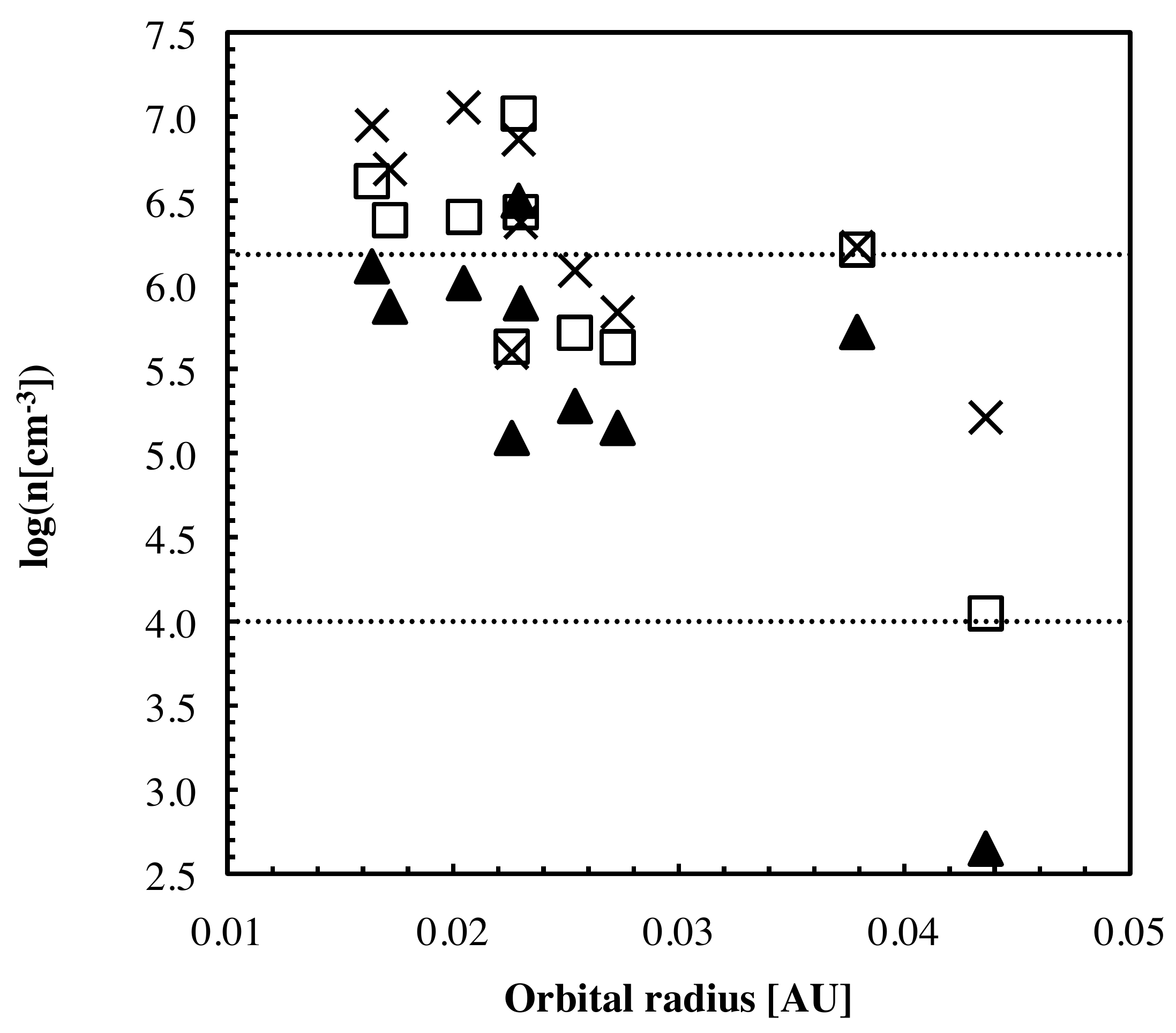}
\caption{Influence of different assumptions on the predicted density at the orbit of each planet in Table~\ref{table2}: for the co-rotating corona with base density that is unscaled (squares) and scaled with $\Omega_*$ (crosses) densities, and for the wind case (triangles). Dashed lines are as in Figure~\ref{dens_all}.}\label{density}
\end{figure}

Assuming that $n_0=n_{0,\odot}$ probably represents effectively a lower limit on the base density because the Sun is a relatively inactive, slowly rotating star. 
For more rapidly rotating, active stars, the coronal density is likely to be higher. To estimate this effect, we scale $n_0$ to the stellar angular velocity $\Omega_*$, such that
\begin{equation}
n_0 = n_{0,\odot} \frac{\Omega_*}{\Omega_\odot}  = n_{0,\odot} \left(   \frac{v \sin(i)}{2~{\rm km~s}^{-1}} \right) \left(\frac{R_\odot}{R_*}\right) .
\end{equation}
As Figure~\ref{density} shows, this has a modest effect on the overall range of predicted densities, but changes the order of planets with the highest predicted coronal densities. 
Thus the three planets with the highest predicted scaled densities are, in order: WASP-18b, WASP-19b, and WASP-12b. Figure~\ref{density} only contains the top 10 planets that may be observable with today's instrumentation, and the special case of CoRoT-11b (Table~\ref{table2}). We note that, although the density predicted for CoRoT-11b (right-most point in Figure~\ref{density}) is just above the threshold limit (lower dashed line), the assumption that the host star coronal base density should scale with $\Omega_*$ leads to a significant increase on the density of the ambient medium surrounding the planet, increasing the potential to detect light curve asymmetries. We note that, in order for a transit to be clearly detected, the depth of the transit should be larger than the instrument's uncertainty. So, for instance, observations of the near-UV transit of CoRoT-7b, which possesses a relatively shallow transit, could be compromised if done with the Cosmic Origins Spectrograph on the HST (used in the case of WASP-12b). Observations of CoRoT-1b may also not result in a positive detection with such instrument, as its host star is about two magnitudes fainter (V band) than WASP-12.

\begin{figure}
	\includegraphics[width=84mm]{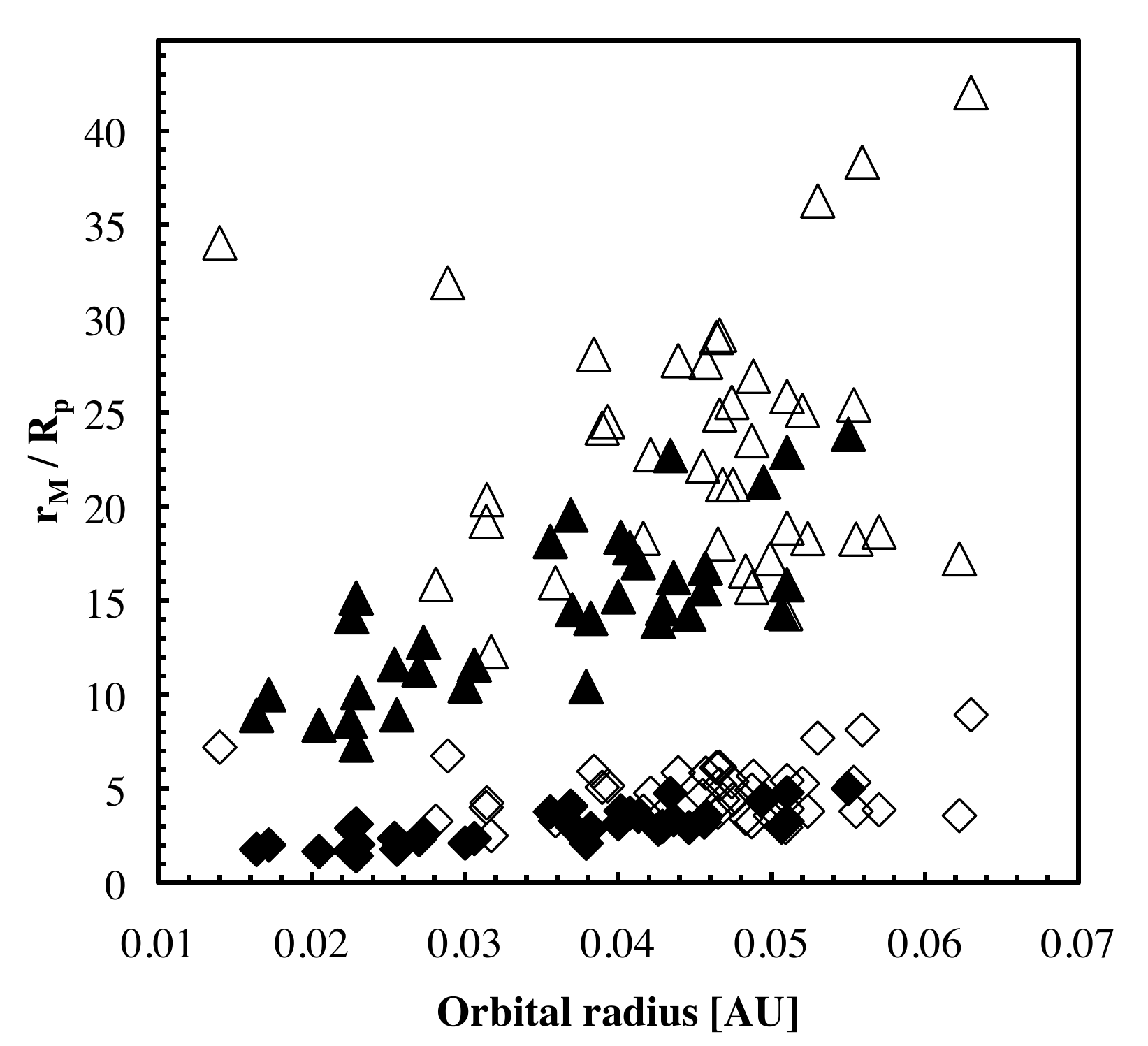}
\caption{Magnetospheric size in units of the planetary radius for a stellar magnetic field strength of $1$~G (triangles) and $100$~G (diamonds). All planets are assumed to have a field of $14$~G. Planets that are surrounded by coronal material denser than $10^4$~cm$^{-3}$ (detection limit) are represented by filled symbols. If we take the stand-off distance between the shock and the planet to be the size of the planet's magnetosphere, planets with large $r_M$ present earlier ingresses (or later egresses) than planets with small $r_M$.}\label{Rmag}
\end{figure}

The stand-off distance of the shock from the planetary surface is also a factor that contributes to the observability of this effect. If we consider that the extent of the planetary magnetosphere $r_M$ can be taken as approximately the observationally derived stand-off distance of the shock, we can calculate the magnetic field intensity of the planet \citepalias{paper1}. In the case where the shock location is determined simply by static pressure balance, we have
\begin{equation}\label{eq.equilibrium}
\frac{m n_c \Delta u^2}{2} + \frac{B_c(R_{\rm orb})^2}{4\pi} + p_c= \frac{B_{p}(r_M)^2}{4\pi} + p_{p} ,
\end{equation}
where $n_c$, $p_c$ and $B_c(R_{\rm orb})$ are the local coronal density, thermal pressure, and magnetic field intensity, and $p_p$ and $B_p (r_M)$ are the planet thermal pressure and magnetic field intensity at $r_M$. In \citetalias{paper1}, we showed for WASP-12b that the dominant terms in Equation~(\ref{eq.equilibrium}) are the magnetic terms. We assume the same is true in our next calculations. Therefore, considering that both the star and the planet magnetic field geometry can be approximated as that of a dipolar field, we have ${B_*}(R_{\rm orb} / R_*)^{-3} \simeq {B_p}{(r_M / R_p)^{-3}}$, where $B_*$ and $B_p$ are the magnetic field strength at the surfaces of the star and the planet, respectively. Hence
\begin{equation}
\frac{r_M }{R_p} =\left(  \frac{B_p}{B_*}  \right)^{1/3} \frac{R_{\rm orb}}{R_*}.
\end{equation}
Therefore, for a given ratio $B_p/B_*$, $r_M$ scales linearly with $R_{\rm orb}$. Figure~\ref{Rmag} shows the effect on $r_M$ due to a change in  $B_*$. The planets are all assumed to have $B_p = 14$~G, similar to that of Jupiter. We obtain values for $r_M$ ranging from $1$ to $40~R_p$ for $B_* = 1$ to $100$~G. These values are comparable to those for the solar system planets. If we take the stand-off distance between the shock and the planet to be $r_M$, planets with large $r_M$ present earlier ingresses (or later egresses) than planets with small $r_M$.

Not all planets may be capable of generating a magnetic field of sufficient strength to sustain a magnetosphere. Table~\ref{table1} shows the values of the minimum ratio $f = (B_p/B_*)_{\rm min}$ required to support a magnetosphere. Note that a planetary magnetic field of about a few percent of that of the host star is enough to support the existence of a planetary magnetosphere. If the kinetic and thermal terms in Equation~\ref{eq.equilibrium} are important, the minimum $B_p$ would even be smaller.

\subsection{Stellar Wind}
If the coronal material is not magnetically confined, it escapes in the form of a wind. By assuming that the host star has an isothermal, thermally-driven wind \citep{parker}, we calculate the wind velocity and density profiles for a few selected planetary systems. Systems that were top-ranked on Table~\ref{table1} but with distances $\gtrsim 1$~kpc where excluded from our calculations, because they are too far away to be observable with the instrumentation available today. One special case (CoRoT-11b) will be discussed in Section~\ref{sec.geometry}. 

The wind radial velocity $u_r$ is derived from the integration of the differential equation
\begin{equation}  
\rho u_r \frac{\partial u_r}{\partial r} = -\frac{\partial p}{\partial r} - \rho \frac{G M_*}{r^2} , 
\end{equation} 
where $\rho = n m$ is the wind mass density, $r$ is the radial distance from the star, $p= n k_B T_{\rm max}$ is the thermal pressure. We assume that the temperature of the wind is the maximum temperature for shock formation given in Table~\ref{table1}\footnote{Note that in the wind scenario, the maximum temperature is actually given by the condition that $\Delta u^2 + u_r^2 = ( u_{\varphi, {\rm wind}} - u_K)^2 + u_r^2 > c_s^2$. Therefore, $T_{\rm max, wind} = {[( u_{\varphi, {\rm wind}} - u_K)^2 + u_r^2]m}/{k_B}$.}. Table~\ref{table2} presents the wind characteristics at $R_{\rm orb}$ for the selected planetary systems. For the top-ranked planets, the $u_r$ ranges from $240$ -- $360$~km~s$^{-1}$ (column 8). The derived local densities (column 11) are obtained considering that the wind base density is the solar value. They are also shown in Figure~\ref{density} (filled triangles) and are $2.5$ -- $3.5$ times smaller than the values derived from the confined corona scenario (shown in column 12 for comparison). The derived mass-loss rates of the host stars (column 14) are at most two orders of magnitude larger than the solar value.

\begin{table*}
\caption{Wind calculation for selected planets. The columns are: (1) the planet name, (2) the stellar mass, (3) and radius, (4) the planet semi-major axis, (5) the maximum temperature necessary for shock formation, (6) the sky-projected stellar rotation velocity, (7) the Keplerian velocity of the planet, (8) the radial velocity of the wind at $R_{\rm orb}$, (9) and (10) the azimuthal velocities of the wind and of the corona at the planetary orbital distance, (11) and (12) the density at the planetary orbital distance for the isothermal wind model and for the coronal model, (13) the angle $\theta$ of the shock normal, and (14) the wind mass-loss rate.\label{table2}}
\begin{center}
\begin{tabular}{lccccccccccccc}
\hline
Planet &  $M_*$ & $R_*$ & $R_{\rm orb}$ & T$_{\rm max}$ & $v \sin(i)$ & $u_{K}$ & $u_{r}$ & $u_{\varphi, {\rm wind}}$  & $u_{\varphi, {\rm cor}} $ & $\log\left[\frac{n}{{\rm cm}^{-3}}\right]$ & $\log\left[\frac{n}{{\rm cm}^{-3}}\right]$ & $\theta$ & $\dot{M}$ \\
Name & ($M_\odot$) & ($R_\odot$) & ($R_*$) & (MK) & (km~s$^{-1}$) & (km~s$^{-1}$) & (km~s$^{-1}$)& (km~s$^{-1}$) & (km~s$^{-1}$)  & wind  & corona & ($^{\rm o}$) & ($M_\odot~{\rm yr}^{-1}$) \\
\scriptsize{(1)} & \scriptsize{(2)} & \scriptsize{(3)} & \scriptsize{(4)} & \scriptsize{(5)} & \scriptsize{(6)} & \scriptsize{(7)} & \scriptsize{(8)}& \scriptsize{(9)} & \scriptsize{(10)}  & \scriptsize{(11)}  & \scriptsize{(12)} & \scriptsize{(13)} & \scriptsize{(14)}\\
\hline \hline	
WASP-12 b	&	$	1.35	$	&	$	1.57	$	&	$	3.14	$	&	$	3.9	$	&	$	2.2	$	&	$	229	$	&	$	358	$	&	$	0.7	$	&	$	6.9	$	&	$	6.50	$	&	$	7.02	$	&	$	58	$	&	$	2.9\times 10^{-12}	$	\\
WASP-19 b	&	$	0.95	$	&	$	0.93	$	&	$	3.79	$	&	$	3.6	$	&	$	4	$	&	$	227	$	&	$	327	$	&	$	1.1	$	&	$	15.2	$	&	$	6.11	$	&	$	6.62	$	&	$	55	$	&	$	5.5\times 10^{-13}	$	\\
WASP-4 b	&	$	0.90	$	&	$	1.15	$	&	$	4.30	$	&	$	2.5	$	&	$	2	$	&	$	186	$	&	$	281	$	&	$	0.5	$	&	$	8.6	$	&	$	5.89	$	&	$	6.43	$	&	$	57	$	&	$	5.7\times 10^{-13}	$	\\
WASP-18 b	&	$	1.28	$	&	$	1.23	$	&	$	3.58	$	&	$	3.1	$	&	$	11	$	&	$	236	$	&	$	260	$	&	$	3.1	$	&	$	39.4	$	&	$	6.01	$	&	$	6.40	$	&	$	48	$	&	$	5.5\times 10^{-13}	$	\\
CoRoT-7 b	&	$	0.93	$	&	$	0.87	$	&	$	4.25	$	&	$	3.3	$	&	$	3.5	$	&	$	219	$	&	$	314	$	&	$	0.8	$	&	$	14.9	$	&	$	5.87	$	&	$	6.38	$	&	$	55	$	&	$	3.4\times 10^{-13}	$	\\
HAT-P-7 b	&	$	1.47	$	&	$	1.84	$	&	$	4.43	$	&	$	2.3	$	&	$	3.8	$	&	$	185	$	&	$	251	$	&	$	0.9	$	&	$	16.8	$	&	$	5.72	$	&	$	6.21	$	&	$	54	$	&	$	9.3\times 10^{-13}	$	\\
CoRoT-1 b	&	$	0.95	$	&	$	1.11	$	&	$	4.92	$	&	$	2.0	$	&	$	5.2	$	&	$	182	$	&	$	217	$	&	$	1.1	$	&	$	25.6	$	&	$	5.28	$	&	$	5.71	$	&	$	50	$	&	$	1.3\times 10^{-13}	$	\\
TrES-3b	&	$	0.92	$	&	$	0.81	$	&	$	5.98	$	&	$	2.6	$	&	$	1.5	$	&	$	190	$	&	$	290	$	&	$	0.3	$	&	$	9.0	$	&	$	5.09	$	&	$	5.63	$	&	$	57	$	&	$	8.9\times 10^{-14}	$	\\
WASP-5 b	&	$	1.02	$	&	$	1.08	$	&	$	5.41	$	&	$	2.1	$	&	$	3.5	$	&	$	182	$	&	$	239	$	&	$	0.6	$	&	$	18.9	$	&	$	5.15	$	&	$	5.63	$	&	$	53	$	&	$	1.2\times 10^{-13}	$	\\
\hline																																																					
CoRoT-11 b	&	$	1.27	$	&	$	1.37	$	&	$	6.84	$	&	$	1.0	$	&	$	40	$	&	$	161	$	&	$	111	$	&	$	5.8	$	&	$	273.7	$	&	$	2.65	$	&	$	4.05	$	&	$	36	$	&	$	4.6\times 10^{-16}	$	\\
\hline
\end{tabular}
\end{center}
\end{table*}

Table~\ref{table2} also presents the azimuthal velocities obtained for the wind and corona scenarios (columns 9 and 10, respectively). Along with the Keplerian velocity $u_K$ of the planet, they provide the condition for shock formation: $|u_K - u_{\varphi, {\rm cor}}| > c_s$ for the corona scenario, and $[|u_K - u_{\varphi, {\rm wind}}|^2+ u_r^2]^{1/2} > c_s$ for the wind scenario. Note that if the corona is confined by the stellar magnetic field, and thus co-rotates with the star, $u_{\varphi, {\rm cor}} = [v \sin(i)/\sin(i)] (R_{\rm orb}/R_*) $. For the wind case, the azimuthal velocity is determined by conservation of angular momentum of the particles leaving the star $u_{\varphi, {\rm wind}} =  [v \sin(i)/\sin(i)] (R_*/R_{\rm orb})$. For the calculations in this paper, we adopt $\sin(i) =1$. If, however, $\sin(i) \ne 1$, then the actual rotation velocity $v$ of the star is larger by a factor $1/\sin(i)$, as well as $u_{\varphi, {\rm cor}}$ and $u_{\varphi, {\rm wind}}$.

The angle that the shock normal makes to the relative azimuthal velocity of the planet is 
\begin{equation}
\theta =  \arctan{\left(\frac{u_r}{|u_K - u_{\varphi, {\rm wind}}|}\right) } .
\end{equation}
Such values are also presented in Table~\ref{table2} (column 13), where we note that $\theta$ ranges between $48$ -- $58^{\rm o}$. As shown in \citetalias{paper1}, $\theta$ increases with the  wind temperature, due to the increase caused in $u_r$ (the thermally-driven wind is more accelerated for large $T$).

\section{Late egress or early ingress?}\label{sec.geometry}
For the ahead-shock case, which occurs in the confined corona scenario, $\theta=0^{\rm o}$, and the shock is always formed perpendicular to the line that connects the centres of the planet and the star. Another limiting case, which was only briefly discussed in \citetalias{paper1}, is the case when the shock trails the planet ($\theta=180^{\rm o}$). In the ``behind-shock'' case, the planet orbits the star beyond the Keplerian co-rotation radius, so that the coronal plasma lags behind the planetary motion. In other words, for a circular orbit, ${\bf u}_K - {\bf u}_{\varphi, {\rm cor}} <0$ or, similarly, $P_*<P_{\rm orb}$, where $P_* = 2 \pi R_*/[v \sin(i)/ \sin(i)]$ is the stellar rotational period. In order to develop a behind-shock, the planet must be in a prograde orbit. 

It is interesting to note that, in the behind-shock limit, the asymmetry in the time of ingress/egress of the light curve will be different from the one observed in WASP-12b, where an ahead-shock is expected to be present \citepalias{paper1}. Because the compressed material trails the planet, in the behind-shock case, the light curve asymmetry occurs at the time of the egress: the planet may have completed the transit through the stellar disk, but the compressed material will still be seen in ``transit'' for an extra amount of time. During this time, if the material is able to absorb stellar radiation in the Mg II lines, for instance, the transit will appear 
to us as a late egress instead of an early ingress, as compared with WASP-12b. Planets with a certain mass escaping rate can also develop cometary tails due to evaporating processes caused by the ionising radiation emitted by the star \citep{ehrenreich2008}. Because a cometary tail around the planet would also produce a late egress, the behind-shock scenario can hence mimic a cometary tail scenario. If CoRoT-7b indeed hosts a cometary tail, as suggested by \citet{mura2010}, then its light curve asymmetry, if observed, will show a late egress. However, we predict the existence of an early ingress caused by the presence of an ahead-shock.

The condition for a behind shock ($P_*<P_{\rm orb}$) was met in $5$ of the $71$ cases where measurements of $v \sin(i)$ exist, namely: WASP-33b, CoRoT-11b, WASP-7b, CoRoT-4b, and CoRoT-6b. WASP-7b, CoRoT-4b, and CoRoT-6b present $10~R_* < R_{\rm orb} < 18~R_*$, which may be too distant to consider that the corona co-rotates with the star. In these three cases, the impacting wind is expected to significantly affect the geometry of the shock, which should be formed at an intermediate angle $\theta$. WASP-33b is an interesting case to study in future works, because its host star is in another mass regime. Therefore, it may have a different internal structure, it should host a different magnetic field structure, which may not confine the stellar material up to the orbit of the planet. As a consequence, the best candidate to host a behind-shock among our sample of planets and, therefore, present a late-egress in the UV light curve asymmetry, is CoRoT-11b. 

CoRoT-11b is an example of a prograde giant gas planet orbiting a rapidly rotating F6-star \citep[$v \sin(i) = 40$~km~s$^{-1}$, ][]{2010arXiv1009.2597G}. If the magnetic field of CoRoT-11 is strong enough to maintain the coronal plasma in co-rotation up to the orbital radius of the planet ($R_{\rm orb} = 6.84~R_*$), then the azimuthal velocity of the corona at the planet orbit is $u_{\varphi, {\rm cor}} \simeq 274$~km~s$^{-1}$. With an orbital period of $P_{\rm orb}=2.99$~d, CoRoT-11b has a Keplerian velocity of $u_K= 161$~km~s$^{-1}$, which leads to a relative azimuthal velocity $\Delta u = |u_K - u_{\varphi, {\rm cor}}| = 113$~km~s$^{-1}$. In order for a shock to form behind the planet, $T<T_{\rm max } \simeq 10^6$~K (Equation~(\ref{eq.tmax})). 

If the coronal plasma is not co-rotating with the star at the orbit of CoRoT-11b, the host star's wind influences shock formation around the planet. By assuming an isothermal wind at $10^6$~K, the shock normal is formed at $\theta \simeq 36^{\rm o}$ (Table~\ref{table2}). However, the low ambient density around the planet ($\sim 25$ times smaller in the wind model compared to the confined corona model), may present some problems for the detection of the absorption during the UV transit. Such a small density is reflected in the sub-solar wind mass-loss rate. It is interesting to note that, although the planet is located in the sub-sonic region of its host star wind, a shock is formed because of the planet's high orbital velocity. 

Therefore, through observations of the planet light curve asymmetry (late egress/early ingress), one can constrain whether CoRoT-11 has a confined corona up to $R_{\rm orb}$ ($\theta=180^{\rm o}$, late egress), or its wind affects shock formation ($\theta\simeq 36^{\rm o}$, early ingress), providing that the coronal temperature allows shock formation.

\section{Conclusion}\label{sec.conclusion}
In this Letter, we determined which planets are prone to develop a surrounding shock that could be observed in asymmetries of planetary light curves. We discussed two possible scenarios for the formation of a shock: (1) when the stellar magnetic field is strong enough to confine the hot coronal plasma out to the planetary orbit and (2) when the stellar magnetic field is unable to confine the plasma, which escapes in a wind. For both cases, we predicted the characteristics of the ambient medium that surrounds the planet, and discussed whether such characteristics presents favourable conditions for the presence of a shock. The planets that were top ranked according to scenario (1) are: WASP-12b, OGLE-TR-56b, WASP-19b, SWEEPS-11, WASP-4b, WASP-18b, CoRoT-7b, CoRoT-14b, HAT-P-7b, OGLE-TR-132b, CoRoT-1b, TrES-3, and  WASP-5b. 

With the observational determination of the stand-off distance between the shock and the planet (which we consider to be the size of the planet's magnetosphere $r_M$), one can infer the planetary magnetic field intensity, provided the stellar magnetic field intensity is known. By assuming that the planet hosts a magnetic field of the same intensity as Jupiter's and the star's magnetic field is in the range between $1$ and $100$~G, we showed that $r_M \simeq 1$ -- $40~R_p$. In general, for the stars in our sample, we show that, to sustain a planetary magnetosphere, a minimum planetary field intensity is required to be only a few percent of the stellar magnetic field.

For the cases where the star is sufficiently magnetised to confine the coronal material, the presence of a shock ahead of the planetary motion can cause an early ingress in the light curve of the planet, while a shock that trails the planet (behind shock) can result in a late egress. We showed that prograde planets orbiting fast-rotating stars can develop a behind-shock. In particular, CoRoT-11b is a promising candidate to host a behind shock.



\def\apj{{ApJ}}                 
\def\apjl{{ApJ}}                
\def\aap{{A\&A}}                

\bsp

\clearpage

\setcounter{table}{0}
\begin{table*}
\caption{Online material. Transiting planetary systems known as of September/2010. Planets are ordered in increasing coronal densities of their host stars, taken as a proxy for the detection of light curves asymmetries. The columns are: (1) the planet name, (2) mass, (3) radius, (4) orbital period, and (5) semi-major axis, (6) the distance to the system, (7) the host star spectral type, (8) mass, and (9) radius, (10) the sky-projected stellar rotation velocity, (11) the maximum temperature required for shock formation; (12) the local density around the planet for the confined corona, and (13) considering the coronal density scales with $\Omega_*$, (14) the size of the planet magnetosphere for a planet with $B_p=14$~G and a star with $B_*=1$~G, (15) the same but for $B_*=100$~G, (16) the minimum planetary magnetic field relative to the stellar one ($f = (B_p/B_*)_{\rm min}$) that is required to sustain a magnetosphere.} 
\begin{center}
\begin{tabular}{lccccccccccccccc}
\hline
Planet&$M_p$&$R_p$&$P_{\rm orb}$&$R_{\rm orb}$&$d$&Spec.&$M_*$&$R_*$&$v \sin(i)$&$T_{\rm max}$&$\log\left[\frac{n}{{\rm cm}^{-3}}\right]$&$\log\left [\frac{n}{{\rm cm}^{-3}}\right]$&	$r_M/R_p$&$r_M/R_p$&$f$	\\
Name&$(M_{\rm J})$&$(R_{\rm J})$&(d)&(AU)&(pc)&Type&$(M_\odot)$&$(R_\odot)$&(km/s)&(MK)&unsc.&scaled&($1$G)&($100$G)&$(\%)$\\
\scriptsize{(1)} & \scriptsize{(2)} & \scriptsize{(3)} & \scriptsize{(4)} & \scriptsize{(5)} & \scriptsize{(6)} & \scriptsize{(7)} & \scriptsize{(8)}& \scriptsize{(9)} & \scriptsize{(10)}  & \scriptsize{(11)}  & \scriptsize{(12)} & \scriptsize{(13)} & \scriptsize{(14)}& \scriptsize{(15)}& \scriptsize{(16)}\\
\hline \hline
WASP-12b	&	$	1.41	$	&	$	1.79	$	&	$	1.09	$	&	$	0.023	$	&	$	267	$	&	G0	&	$	1.35	$	&	$	1.57	$	&	$	2.2	$	&	$	3.96	$	&	$	7.02	$	&	$	6.86	$	&	$	7.5	$	&	$	1.6	$	&	$	3.2	$	\\
OGLE-TR-56b	&	$	1.30	$	&	$	1.20	$	&	$	1.21	$	&	$	0.023	$	&	$	1500	$	&	G	&	$	1.17	$	&	$	1.32	$	&	$	3.2	$	&	$	3.32	$	&	$	6.71	$	&	$	6.80	$	&	$	8.8	$	&	$	1.9	$	&	$	2.0	$	\\
WASP-19b	&	$	1.15	$	&	$	1.31	$	&	$	0.79	$	&	$	0.016	$	&	$	-	$	&	G8V	&	$	0.95	$	&	$	0.93	$	&	$	4	$	&	$	3.61	$	&	$	6.62	$	&	$	6.95	$	&	$	9.1	$	&	$	2.0	$	&	$	1.8	$	\\
SWEEPS-11 	&	$	9.70	$	&	$	1.13	$	&	$	1.80	$	&	$	0.030	$	&	$	8500	$	&	$-$	&	$	1.10	$	&	$	1.45	$	&	$	-	$	&	$	2.62	$	&	$	6.51	$	&	$	-	$	&	$	10.7	$	&	$	2.3	$	&	$	1.1	$	\\
WASP-4b	&	$	1.12	$	&	$	1.42	$	&	$	1.34	$	&	$	0.023	$	&	$	300	$	&	G7V	&	$	0.90	$	&	$	1.15	$	&	$	2	$	&	$	2.55	$	&	$	6.43	$	&	$	6.37	$	&	$	10.3	$	&	$	2.2	$	&	$	1.3	$	\\
WASP-18b	&	$	10.43	$	&	$	1.17	$	&	$	0.94	$	&	$	0.020	$	&	$	100	$	&	F9	&	$	1.28	$	&	$	1.23	$	&	$	11	$	&	$	3.11	$	&	$	6.40	$	&	$	7.05	$	&	$	8.6	$	&	$	1.9	$	&	$	2.2	$	\\
CoRoT-7b	&	$	0.02	$	&	$	0.15	$	&	$	0.85	$	&	$	0.017	$	&	$	150	$	&	K0V	&	$	0.93	$	&	$	0.87	$	&	$	3.5	$	&	$	3.36	$	&	$	6.38	$	&	$	6.69	$	&	$	10.2	$	&	$	2.2	$	&	$	1.3	$	\\
CoRoT-14b	&	$	7.60	$	&	$	1.09	$	&	$	1.51	$	&	$	0.027	$	&	$	1340	$	&	F9V	&	$	1.13	$	&	$	1.21	$	&	$	-	$	&	$	2.99	$	&	$	6.36	$	&	$	-	$	&	$	11.5	$	&	$	2.5	$	&	$	0.91	$	\\
HAT-P-7b	&	$	1.80	$	&	$	1.42	$	&	$	2.20	$	&	$	0.038	$	&	$	320	$	&	$-$	&	$	1.47	$	&	$	1.84	$	&	$	3.8	$	&	$	2.29	$	&	$	6.21	$	&	$	6.22	$	&	$	10.6	$	&	$	2.3	$	&	$	1.2	$	\\
OGLE-TR-132b	&	$	1.17	$	&	$	1.25	$	&	$	1.69	$	&	$	0.031	$	&	$	1500	$	&	F	&	$	1.26	$	&	$	1.34	$	&	$	5	$	&	$	2.24	$	&	$	5.78	$	&	$	6.05	$	&	$	11.8	$	&	$	2.5	$	&	$	0.84	$	\\
CoRoT-1b	&	$	1.03	$	&	$	1.49	$	&	$	1.51	$	&	$	0.025	$	&	$	460	$	&	G0V	&	$	0.95	$	&	$	1.11	$	&	$	5.2	$	&	$	1.98	$	&	$	5.71	$	&	$	6.08	$	&	$	11.8	$	&	$	2.6	$	&	$	0.84	$	\\
TrES-3 	&	$	1.91	$	&	$	1.31	$	&	$	1.31	$	&	$	0.023	$	&	$	-	$	&	G	&	$	0.92	$	&	$	0.81	$	&	$	1.5	$	&	$	2.65	$	&	$	5.63	$	&	$	5.60	$	&	$	14.3	$	&	$	3.1	$	&	$	0.47	$	\\
WASP-5b	&	$	1.64	$	&	$	1.17	$	&	$	1.63	$	&	$	0.027	$	&	$	297	$	&	G4V	&	$	1.02	$	&	$	1.08	$	&	$	3.5	$	&	$	2.15	$	&	$	5.63	$	&	$	5.84	$	&	$	13.0	$	&	$	2.8	$	&	$	0.63	$	\\
OGLE-TR-211b	&	$	0.75	$	&	$	1.26	$	&	$	3.68	$	&	$	0.051	$	&	$	-	$	&	$-$	&	$	1.33	$	&	$	1.64	$	&	$	-	$	&	$	1.86	$	&	$	5.54	$	&	$	-	$	&	$	16.0	$	&	$	3.5	$	&	$	0.33	$	\\
HAT-P-13b	&	$	0.85	$	&	$	1.28	$	&	$	2.92	$	&	$	0.043	$	&	$	214	$	&	G4	&	$	1.22	$	&	$	1.56	$	&	$	2.9	$	&	$	1.63	$	&	$	5.37	$	&	$	5.33	$	&	$	14.1	$	&	$	3.0	$	&	$	0.49	$	\\
WASP-14b	&	$	7.73	$	&	$	1.26	$	&	$	2.24	$	&	$	0.037	$	&	$	160	$	&	F5V	&	$	1.32	$	&	$	1.30	$	&	$	2.8	$	&	$	2.08	$	&	$	5.29	$	&	$	5.32	$	&	$	14.7	$	&	$	3.2	$	&	$	0.43	$	\\
HAT-P-24b	&	$	0.69	$	&	$	1.24	$	&	$	3.36	$	&	$	0.047	$	&	$	306	$	&	$-$	&	$	1.19	$	&	$	1.32	$	&	$	-	$	&	$	1.83	$	&	$	5.15	$	&	$	-	$	&	$	18.2	$	&	$	3.9	$	&	$	0.23	$	\\
WASP-26b	&	$	1.02	$	&	$	1.32	$	&	$	2.76	$	&	$	0.040	$	&	$	250	$	&	G0	&	$	1.12	$	&	$	1.34	$	&	$	2.4	$	&	$	1.63	$	&	$	5.13	$	&	$	5.08	$	&	$	15.4	$	&	$	3.3	$	&	$	0.38	$	\\
CoRoT-12b	&	$	0.92	$	&	$	1.44	$	&	$	2.83	$	&	$	0.040	$	&	$	1150	$	&	G2V	&	$	1.08	$	&	$	1.12	$	&	$	-	$	&	$	1.92	$	&	$	5.09	$	&	$	-	$	&	$	18.6	$	&	$	4.0	$	&	$	0.22	$	\\
Kepler-4b	&	$	0.08	$	&	$	0.36	$	&	$	3.21	$	&	$	0.046	$	&	$	550	$	&	G0	&	$	1.22	$	&	$	1.49	$	&	$	2.2	$	&	$	1.57	$	&	$	5.06	$	&	$	4.93	$	&	$	15.8	$	&	$	3.4	$	&	$	0.35	$	\\
WASP-33b	&	$	4.1	$	&	$	1.5	$	&	$	1.22	$	&	$	0.03	$	&	$	116	$	&	A5	&	$	1.50	$	&	$	1.44	$	&	$	90	$	&	$	1.03	$	&	$	4.92	$	&	$	6.42	$	&	$	9.1	$	&	$	2.0	$	&	$	1.8	$	\\
WASP-1b	&	$	0.86	$	&	$	1.48	$	&	$	2.52	$	&	$	0.038	$	&	$	-	$	&	F7V	&	$	1.24	$	&	$	1.38	$	&	$	5	$	&	$	1.58	$	&	$	4.87	$	&	$	5.13	$	&	$	14.3	$	&	$	3.1	$	&	$	0.48	$	\\
Kepler-5b	&	$	2.11	$	&	$	1.43	$	&	$	3.55	$	&	$	0.051	$	&	$	-	$	&	$-$	&	$	1.37	$	&	$	1.79	$	&	$	4.8	$	&	$	1.28	$	&	$	4.69	$	&	$	4.82	$	&	$	14.6	$	&	$	3.2	$	&	$	0.45	$	\\
HAT-P-4b	&	$	0.68	$	&	$	1.27	$	&	$	3.06	$	&	$	0.045	$	&	$	310	$	&	F	&	$	1.26	$	&	$	1.59	$	&	$	5.5	$	&	$	1.26	$	&	$	4.54	$	&	$	4.78	$	&	$	14.5	$	&	$	3.1	$	&	$	0.46	$	\\
OGLE-TR-113b	&	$	1.24	$	&	$	1.11	$	&	$	1.43	$	&	$	0.023	$	&	$	1500	$	&	K	&	$	0.78	$	&	$	0.77	$	&	$	5	$	&	$	1.62	$	&	$	4.51	$	&	$	5.03	$	&	$	15.3	$	&	$	3.3	$	&	$	0.38	$	\\
Kepler-6b	&	$	0.67	$	&	$	1.32	$	&	$	3.23	$	&	$	0.046	$	&	$	-	$	&	$-$	&	$	1.21	$	&	$	1.39	$	&	$	3	$	&	$	1.41	$	&	$	4.48	$	&	$	4.52	$	&	$	16.9	$	&	$	3.7	$	&	$	0.28	$	\\
TrES-2 	&	$	1.25	$	&	$	1.26	$	&	$	2.47	$	&	$	0.036	$	&	$	220	$	&	G0V	&	$	0.98	$	&	$	1.00	$	&	$	2	$	&	$	1.60	$	&	$	4.48	$	&	$	4.48	$	&	$	18.3	$	&	$	4.0	$	&	$	0.22	$	\\
XO-2b	&	$	0.57	$	&	$	0.97	$	&	$	2.62	$	&	$	0.037	$	&	$	149	$	&	K0V	&	$	0.98	$	&	$	0.96	$	&	$	1.3	$	&	$	1.64	$	&	$	4.39	$	&	$	4.22	$	&	$	19.8	$	&	$	4.3	$	&	$	0.18	$	\\
HAT-P-5b	&	$	1.06	$	&	$	1.26	$	&	$	2.79	$	&	$	0.041	$	&	$	340	$	&	$-$	&	$	1.16	$	&	$	1.17	$	&	$	2.6	$	&	$	1.57	$	&	$	4.35	$	&	$	4.40	$	&	$	18.0	$	&	$	3.9	$	&	$	0.24	$	\\
HD 149026b	&	$	0.36	$	&	$	0.61	$	&	$	2.88	$	&	$	0.043	$	&	$	78.9	$	&	G0 IV	&	$	1.30	$	&	$	1.50	$	&	$	6	$	&	$	1.30	$	&	$	4.31	$	&	$	4.61	$	&	$	14.8	$	&	$	3.2	$	&	$	0.43	$	\\
WASP-37b	&	$	1.70	$	&	$	1.14	$	&	$	3.58	$	&	$	0.043	$	&	$	338	$	&	G2	&	$	0.85	$	&	$	0.98	$	&	$	-	$	&	$	1.40	$	&	$	4.30	$	&	$	-	$	&	$	22.9	$	&	$	5.0	$	&	$	0.11	$	\\
OGLE-TR-182b	&	$	1.06	$	&	$	1.47	$	&	$	3.98	$	&	$	0.051	$	&	$	-	$	&	$-$	&	$	1.14	$	&	$	1.14	$	&	$	-	$	&	$	1.60	$	&	$	4.27	$	&	$	-	$	&	$	23.1	$	&	$	5.0	$	&	$	0.11	$	\\
HAT-P-16b	&	$	4.19	$	&	$	1.29	$	&	$	2.78	$	&	$	0.041	$	&	$	235	$	&	F8	&	$	1.22	$	&	$	1.24	$	&	$	3.5	$	&	$	1.51	$	&	$	4.27	$	&	$	4.42	$	&	$	17.2	$	&	$	3.7	$	&	$	0.27	$	\\
SWEEPS-04 	&	$	3.80	$	&	$	0.81	$	&	$	4.20	$	&	$	0.055	$	&	$	8500	$	&	$-$	&	$	1.24	$	&	$	1.18	$	&	$	-	$	&	$	1.61	$	&	$	4.10	$	&	$	-	$	&	$	24.1	$	&	$	5.2	$	&	$	0.10	$	\\
CoRoT-11b	&	$	2.33	$	&	$	1.43	$	&	$	2.99	$	&	$	0.044	$	&	$	560	$	&	F6V	&	$	1.27	$	&	$	1.37	$	&	$	40^a	$	&	$	1.01	$	&	$	4.05	$	&	$	5.21	$	&	$	16.4	$	&	$	3.6	$	&	$	0.31	$	\\
CoRoT-5b	&	$	0.47	$	&	$	1.39	$	&	$	4.04	$	&	$	0.049	$	&	$	400	$	&	F9V	&	$	1.00	$	&	$	1.19	$	&	$	1	$	&	$	1.26	$	&	$	4.05	$	&	$	3.67	$	&	$	21.5	$	&	$	4.7	$	&	$	0.14	$	\\
HAT-P-25b	&	$	0.57	$	&	$	1.19	$	&	$	3.65	$	&	$	0.047	$	&	$	297	$	&	G5	&	$	1.01	$	&	$	0.96	$	&	$	-	$	&	$	1.55	$	&	$	3.92	$	&	$	-	$	&	$	25.1	$	&	$	5.4	$	&	$	0.09	$	\\
WASP-15b	&	$	0.54	$	&	$	1.43	$	&	$	3.75	$	&	$	0.050	$	&	$	308	$	&	F5	&	$	1.18	$	&	$	1.48	$	&	$	4	$	&	$	1.08	$	&	$	3.78	$	&	$	3.91	$	&	$	17.4	$	&	$	3.8	$	&	$	0.26	$	\\
HAT-P-3b	&	$	0.60	$	&	$	0.89	$	&	$	2.90	$	&	$	0.039	$	&	$	140	$	&	K	&	$	0.94	$	&	$	0.82	$	&	$	0.5	$	&	$	1.60	$	&	$	3.75	$	&	$	3.23	$	&	$	24.4	$	&	$	5.3	$	&	$	0.10	$	\\
CoRoT-13b	&	$	1.31	$	&	$	0.89	$	&	$	4.04	$	&	$	0.051	$	&	$	1310	$	&	G0V	&	$	1.09	$	&	$	1.01	$	&	$	-	$	&	$	1.53	$	&	$	3.74	$	&	$	-	$	&	$	26.1	$	&	$	5.6	$	&	$	0.08	$	\\
XO-5b	&	$	1.08	$	&	$	1.09	$	&	$	4.19	$	&	$	0.049	$	&	$	255	$	&	G8V	&	$	0.88	$	&	$	1.06	$	&	$	0.7	$	&	$	1.15	$	&	$	3.71	$	&	$	3.23	$	&	$	23.7	$	&	$	5.1	$	&	$	0.10	$	\\
Kepler-7b	&	$	0.43	$	&	$	1.48	$	&	$	4.89	$	&	$	0.062	$	&	$	-	$	&	$-$	&	$	1.35	$	&	$	1.84	$	&	$	4.2	$	&	$	0.94	$	&	$	3.58	$	&	$	3.63	$	&	$	17.4	$	&	$	3.8	$	&	$	0.26	$	\\
WASP-3b	&	$	2.06	$	&	$	1.45	$	&	$	1.85	$	&	$	0.032	$	&	$	223	$	&	F7V	&	$	1.24	$	&	$	1.31	$	&	$	13.4	$	&	$	1.10	$	&	$	3.45	$	&	$	4.16	$	&	$	12.5	$	&	$	2.7	$	&	$	0.71	$	\\
WASP-24b	&	$	1.03	$	&	$	1.10	$	&	$	2.34	$	&	$	0.036	$	&	$	330	$	&	F8-9	&	$	1.13	$	&	$	1.15	$	&	$	6.96	$	&	$	1.17	$	&	$	3.26	$	&	$	3.74	$	&	$	16.2	$	&	$	3.5	$	&	$	0.33	$	\\
TrES-1 	&	$	0.76	$	&	$	1.10	$	&	$	3.03	$	&	$	0.039	$	&	$	157	$	&	K0V	&	$	0.87	$	&	$	0.82	$	&	$	1.08	$	&	$	1.34	$	&	$	3.26	$	&	$	3.08	$	&	$	24.7	$	&	$	5.3	$	&	$	0.09	$	\\
TrES-4 	&	$	0.88	$	&	$	1.81	$	&	$	3.55	$	&	$	0.051	$	&	$	440	$	&	F	&	$	1.38	$	&	$	1.81	$	&	$	8.5	$	&	$	0.87	$	&	$	3.19	$	&	$	3.56	$	&	$	14.5	$	&	$	3.1	$	&	$	0.45	$	\\
Lupus-TR-3b	&	$	0.81	$	&	$	0.89	$	&	$	3.91	$	&	$	0.046	$	&	$	-	$	&	K1V	&	$	0.87	$	&	$	0.82	$	&	$	-	$	&	$	1.34	$	&	$	3.17	$	&	$	-	$	&	$	29.2	$	&	$	6.3	$	&	$	0.06	$	\\
HAT-P-19b	&	$	0.29	$	&	$	1.13	$	&	$	4.01	$	&	$	0.047	$	&	$	215	$	&	K	&	$	0.84	$	&	$	0.82	$	&	$	-	$	&	$	1.29	$	&	$	3.15	$	&	$	-	$	&	$	29.3	$	&	$	6.3	$	&	$	0.05	$	\\
WASP-28b	&	$	0.91	$	&	$	1.12	$	&	$	3.41	$	&	$	0.046	$	&	$	334	$	&	F8-G0	&	$	1.08	$	&	$	1.05	$	&	$	2.2	$	&	$	1.25	$	&	$	3.14	$	&	$	3.16	$	&	$	22.4	$	&	$	4.8	$	&	$	0.12	$	\\
HD 189733b	&	$	1.15	$	&	$	1.15	$	&	$	2.22	$	&	$	0.031	$	&	$	19.3	$	&	K1-K2	&	$	0.80	$	&	$	0.79	$	&	$	3.32	$	&	$	1.20	$	&	$	3.04	$	&	$	3.36	$	&	$	20.6	$	&	$	4.4	$	&	$	0.16	$	\\
${{\rm WASP-11}\atop {\rm HAT-P-10b}}$	&	$	0.46	$	&	$	1.05	$	&	$	3.72	$	&	$	0.044	$	&	$	125	$	&	K3V	&	$	0.82	$	&	$	0.81	$	&	$	0.5	$	&	$	1.22	$	&	$	2.95	$	&	$	2.44	$	&	$	28.0	$	&	$	6.0	$	&	$	0.06	$	\\
HAT-P-12b	&	$	0.21	$	&	$	0.96	$	&	$	3.21	$	&	$	0.038	$	&	$	142.5	$	&	$-$	&	$	0.73	$	&	$	0.70	$	&	$	0.5	$	&	$	1.24	$	&	$	2.88	$	&	$	2.43	$	&	$	28.3	$	&	$	6.1	$	&	$	0.06	$	\\
WASP-21b	&	$	0.30	$	&	$	1.07	$	&	$	4.32	$	&	$	0.052	$	&	$	230	$	&	G3V	&	$	1.01	$	&	$	1.06	$	&	$	1.5	$	&	$	1.07	$	&	$	2.67	$	&	$	2.52	$	&	$	25.3	$	&	$	5.5	$	&	$	0.09	$	\\

\hline
\end{tabular}
\end{center}
$^a$ Gandolfi et al. (2010)
\end{table*}

\begin{table*}
\contcaption{Online material.}
\begin{center}
\begin{tabular}{lccccccccccccccc}
\hline
Planet&$M_p$&$R_p$&$P_{\rm orb}$&$R_{\rm orb}$&$d$&Spec.&$M_*$&$R_*$&$v \sin(i)$&$T_{\rm max}$&$\log\left[\frac{n}{{\rm cm}^{-3}}\right]$&$\log\left [\frac{n}{{\rm cm}^{-3}}\right]$&	$r_M/R_p$&$r_M/R_p$&$f$	\\
Name&$(M_{\rm J})$&$(R_{\rm J})$&(d)&(AU)&(pc)&Type&$(M_\odot)$&$(R_\odot)$&(km/s)&(MK)&unsc.&scaled&($1$G)&($100$G)&$(\%)$\\
\scriptsize{(1)} & \scriptsize{(2)} & \scriptsize{(3)} & \scriptsize{(4)} & \scriptsize{(5)} & \scriptsize{(6)} & \scriptsize{(7)} & \scriptsize{(8)}& \scriptsize{(9)} & \scriptsize{(10)}  & \scriptsize{(11)}  & \scriptsize{(12)} & \scriptsize{(13)} & \scriptsize{(14)}& \scriptsize{(15)}& \scriptsize{(16)}\\  
\hline \hline
XO-1b	&	$	0.90	$	&	$	1.18	$	&	$	3.94	$	&	$	0.049	$	&	$	200	$	&	G1V	&	$	1.00	$	&	$	0.93	$	&	$	1.11	$	&	$	1.21	$	&	$	2.59	$	&	$	2.37	$	&	$	27.1	$	&	$	5.9	$	&	$	0.07	$	\\
WASP-22b	&	$	0.56	$	&	$	1.12	$	&	$	3.53	$	&	$	0.05	$	&	$	300	$	&	$-$	&	$	1.10	$	&	$	1.13	$	&	$	3.5	$	&	$	1.04	$	&	$	2.47	$	&	$	2.66	$	&	$	21.4	$	&	$	4.6	$	&	$	0.14	$	\\
WASP-2b	&	$	0.85	$	&	$	1.04	$	&	$	2.15	$	&	$	0.031	$	&	$	144	$	&	K1V	&	$	0.84	$	&	$	0.83	$	&	$	5	$	&	$	1.04	$	&	$	2.41	$	&	$	2.88	$	&	$	19.4	$	&	$	4.2	$	&	$	0.19	$	\\
WASP-16b	&	$	0.86	$	&	$	1.01	$	&	$	3.12	$	&	$	0.042	$	&	$	-	$	&	G3V	&	$	1.02	$	&	$	0.95	$	&	$	3	$	&	$	1.12	$	&	$	2.30	$	&	$	2.50	$	&	$	23.0	$	&	$	5.0	$	&	$	0.11	$	\\
WASP-29b	&	$	0.24	$	&	$	0.79	$	&	$	3.92	$	&	$	0.046	$	&	$	80	$	&	K4V	&	$	0.82	$	&	$	0.85	$	&	$	1.5	$	&	$	0.96	$	&	$	1.84	$	&	$	1.79	$	&	$	27.9	$	&	$	6.0	$	&	$	0.06	$	\\
HAT-P-18b	&	$	0.20	$	&	$	1.00	$	&	$	5.51	$	&	$	0.056	$	&	$	166	$	&	K	&	$	0.77	$	&	$	0.75	$	&	$	-	$	&	$	0.98	$	&	$	1.49	$	&	$	-	$	&	$	38.5	$	&	$	8.3	$	&	$	0.02	$	\\
OGLE-TR-10b	&	$	0.68	$	&	$	1.72	$	&	$	3.10	$	&	$	0.042	$	&	$	1500	$	&	G or K	&	$	1.18	$	&	$	1.16	$	&	$	7	$	&	$	0.88	$	&	$	1.41	$	&	$	1.89	$	&	$	18.5	$	&	$	4.0	$	&	$	0.22	$	\\
HD 209458b	&	$	0.64	$	&	$	1.38	$	&	$	3.52	$	&	$	0.047	$	&	$	47	$	&	G0 V	&	$	1.00	$	&	$	1.15	$	&	$	4.7	$	&	$	0.73	$	&	$	0.96	$	&	$	1.28	$	&	$	21.4	$	&	$	4.6	$	&	$	0.14	$	\\
WASP-25b	&	$	0.58	$	&	$	1.26	$	&	$	3.76	$	&	$	0.047	$	&	$	169	$	&	G4	&	$	1.00	$	&	$	0.95	$	&	$	3	$	&	$	0.88	$	&	$	0.85	$	&	$	1.04	$	&	$	25.7	$	&	$	5.6	$	&	$	0.08	$	\\
CoRoT-8b	&	$	0.22	$	&	$	0.57	$	&	$	6.21	$	&	$	0.06	$	&	$	380	$	&	K1V	&	$	0.88	$	&	$	0.77	$	&	$	-	$	&	$	1	$	&	$	0.83	$	&	$	-	$	&	$	42.2	$	&	$	9.1	$	&	$	0.02	$	\\
HAT-P-15b	&	$	1.95	$	&	$	1.07	$	&	$	10.86	$	&	$	0.096	$	&	$	190	$	&	G5	&	$	1.01	$	&	$	1.08	$	&	$	-	$	&	$	0.75	$	&	$	0.13	$	&	$	-	$	&	$	46.1	$	&	$	10.0	$	&	$	0.01	$	\\
CoRoT-2b	&	$	3.31	$	&	$	1.47	$	&	$	1.74	$	&	$	0.03	$	&	$	300	$	&	G7V	&	$	0.97	$	&	$	0.90	$	&	$	11.85	$	&	$	0.74	$	&	$	-0.05	$	&	$	0.76	$	&	$	16.1	$	&	$	3.5	$	&	$	0.33	$	\\
HAT-P-1b	&	$	0.52	$	&	$	1.22	$	&	$	4.47	$	&	$	0.055	$	&	$	139	$	&	GOV	&	$	1.13	$	&	$	1.12	$	&	$	3.75	$	&	$	0.73	$	&	$	-0.40	$	&	$	-0.17	$	&	$	25.6	$	&	$	5.5	$	&	$	0.08	$	\\
HAT-P-8b	&	$	1.52	$	&	$	1.50	$	&	$	3.08	$	&	$	0.049	$	&	$	230	$	&	$-$	&	$	1.28	$	&	$	1.58	$	&	$	11.5	$	&	$	0.48	$	&	$	-1.42	$	&	$	-0.85	$	&	$	15.9	$	&	$	3.4	$	&	$	0.34	$	\\
HAT-P-11b	&	$	0.08	$	&	$	0.45	$	&	$	4.89	$	&	$	0.05	$	&	$	38	$	&	K4	&	$	0.81	$	&	$	0.75	$	&	$	1.5	$	&	$	0.71	$	&	$	-1.46	$	&	$	-1.46	$	&	$	36.5	$	&	$	7.9	$	&	$	0.03	$	\\
HAT-P-6b	&	$	1.06	$	&	$	1.33	$	&	$	3.85	$	&	$	0.052	$	&	$	200	$	&	F	&	$	1.29	$	&	$	1.46	$	&	$	8.7	$	&	$	0.53	$	&	$	-1.51	$	&	$	-1.04	$	&	$	18.5	$	&	$	4.0	$	&	$	0.22	$	\\
Kepler-8b	&	$	0.6	$	&	$	1.42	$	&	$	3.52	$	&	$	0.05	$	&	$	1330	$	&	$-$	&	$	1.21	$	&	$	1.49	$	&	$	10.5	$	&	$	0.47	$	&	$	-1.75	$	&	$	-1.20	$	&	$	16.8	$	&	$	3.6	$	&	$	0.29	$	\\
XO-4b	&	$	1.72	$	&	$	1.34	$	&	$	4.13	$	&	$	0.06	$	&	$	293	$	&	F5V	&	$	1.32	$	&	$	1.55	$	&	$	8.8	$	&	$	0.49	$	&	$	-1.94	$	&	$	-1.49	$	&	$	18.5	$	&	$	4.0	$	&	$	0.22	$	\\
GJ 436b	&	$	0.07	$	&	$	0.37	$	&	$	2.64	$	&	$	0.029	$	&	$	10.2	$	&	M2.5	&	$	0.45	$	&	$	0.46	$	&	$	2.4	$	&	$	0.59	$	&	$	-2.05	$	&	$	-1.64	$	&	$	32.1	$	&	$	6.9	$	&	$	0.04	$	\\
GJ 1214b	&	$	0.02	$	&	$	0.24	$	&	$	1.58	$	&	$	0.014	$	&	$	13	$	&	$-$	&	$	0.16	$	&	$	0.21	$	&	$	2	$	&	$	0.41	$	&	$	-3.18	$	&	$	-2.51	$	&	$	34.2	$	&	$	7.4	$	&	$	0.03	$	\\
WASP-17b	&	$	0.49	$	&	$	1.74	$	&	$	3.74	$	&	$	0.051	$	&	$	-	$	&	F6	&	$	1.20	$	&	$	1.38	$	&	$	9	$	&	$	0.43	$	&	$	-3.48	$	&	$	-2.97	$	&	$	19.1	$	&	$	4.1	$	&	$	0.20	$	\\
HAT-P-14b	&	$	2.2	$	&	$	1.2	$	&	$	4.63	$	&	$	0.06	$	&	$	205	$	&	F	&	$	1.39	$	&	$	1.47	$	&	$	8.4	$	&	$	0.41	$	&	$	-5.42	$	&	$	-4.96	$	&	$	20.9	$	&	$	4.5	$	&	$	0.15	$	\\
WASP-10b	&	$	3.06	$	&	$	1.08	$	&	$	3.09	$	&	$	0.037	$	&	$	90	$	&	K5	&	$	0.71	$	&	$	0.78	$	&	$	6	$	&	$	0.39	$	&	$	-5.83	$	&	$	-5.25	$	&	$	24.4	$	&	$	5.3	$	&	$	0.09	$	\\
OGLE-TR-111b	&	$	0.54	$	&	$	1.08	$	&	$	4.01	$	&	$	0.047	$	&	$	1500	$	&	G or K	&	$	0.82	$	&	$	0.83	$	&	$	5	$	&	$	0.33	$	&	$	-10.13	$	&	$	-9.65	$	&	$	29.2	$	&	$	6.3	$	&	$	0.06	$	\\
WASP-7b	&	$	0.96	$	&	$	0.92	$	&	$	4.95	$	&	$	0.062	$	&	$	140	$	&	F5V	&	$	1.28	$	&	$	1.24	$	&	$	17	$	&	$	0.17	$	&	$	-24.53	$	&	$	-23.70	$	&	$	25.8	$	&	$	5.6	$	&	$	0.08	$	\\
HAT-P-9b	&	$	0.78	$	&	$	1.40	$	&	$	3.92	$	&	$	0.053	$	&	$	480	$	&	F	&	$	1.28	$	&	$	1.32	$	&	$	11.9	$	&	$	0.16	$	&	$	-27.30	$	&	$	-26.64	$	&	$	20.7	$	&	$	4.5	$	&	$	0.16	$	\\
HAT-P-23b	&	$	2.09	$	&	$	1.37	$	&	$	1.21	$	&	$	0.023	$	&	$	-	$	&	$-$	&	$	-	$	&	$	-	$	&	$	-	$	&	$	-	$	&	$	-	$	&	$	-	$	&	$	-	$	&	$	-	$	&	$	-	$	\\
HAT-P-22b	&	$	2.15	$	&	$	1.08	$	&	$	3.21	$	&	$	0.041	$	&	$	-	$	&	$-$	&	$	-	$	&	$	-	$	&	$	-	$	&	$	-	$	&	$	-	$	&	$	-	$	&	$	-	$	&	$	-	$	&	$	-	$	\\
WASP-6b	&	$	0.50	$	&	$	1.22	$	&	$	3.36	$	&	$	0.042	$	&	$	307	$	&	G8	&	$	-	$	&	$	-	$	&	$	1.4	$	&	$	-	$	&	$	-	$	&	$	-	$	&	$	-	$	&	$	-	$	&	$	0.00	$	\\
XO-3b	&	$	11.79	$	&	$	1.22	$	&	$	3.19	$	&	$	0.045	$	&	$	260	$	&	F5V	&	$	1.21	$	&	$	1.38	$	&	$	18.54	$	&	$	0.04	$	&	$	-	$	&	$	-	$	&	$	17.0	$	&	$	3.7	$	&	$	0.28	$	\\
WASP-36b	&	$	2.40	$	&	$	1.40	$	&	$	1.50	$	&	$	-	$	&	$	-	$	&	$-$	&	$	-	$	&	$	-	$	&	$	-	$	&	$	-	$	&	$	-	$	&	$	-	$	&	$	-	$	&	$	-	$	&	$	-	$	\\
HAT-P-20b	&	$	7.25	$	&	$	0.87	$	&	$	2.88	$	&	$	0.036	$	&	$	-	$	&	$-$	&	$	-	$	&	$	-	$	&	$	-	$	&	$	-	$	&	$	-	$	&	$	-	$	&	$	-	$	&	$	-	$	&	$	-	$	\\
HAT-P-21b	&	$	4.06	$	&	$	1.02	$	&	$	4.12	$	&	$	0.049	$	&	$	-	$	&	$-$	&	$	-	$	&	$	-	$	&	$	-	$	&	$	-	$	&	$	-	$	&	$	-	$	&	$	-	$	&	$	-	$	&	$	-	$	\\
CoRoT-3b	&	$	21.66	$	&	$	1.01	$	&	$	4.26	$	&	$	0.057	$	&	$	680	$	&	F3V	&	$	1.37	$	&	$	1.56	$	&	$	17	$	&	$	0.01	$	&	$	-	$	&	$	-	$	&	$	18.9	$	&	$	4.1	$	&	$	0.21	$	\\
CoRoT-4b	&	$	0.72	$	&	$	1.19	$	&	$	9.20	$	&	$	0.090	$	&	$	-	$	&	F8V	&	$	1.10	$	&	$	1.15	$	&	$	6.4	$	&	$	0.00	$	&	$	-	$	&	$	-	$	&	$	40.4	$	&	$	8.7	$	&	$	0.02	$	\\
CoRoT-6b	&	$	2.96	$	&	$	1.17	$	&	$	8.89	$	&	$	0.086	$	&	$	-	$	&	F9V	&	$	1.06	$	&	$	1.03	$	&	$	7.5	$	&	$	0.07	$	&	$	-	$	&	$	-	$	&	$	43.0	$	&	$	9.3	$	&	$	0.02	$	\\
OGLE2-TR-L9b	&	$	4.34	$	&	$	1.61	$	&	$	2.49	$	&	$	-	$	&	$	900	$	&	F3	&	$	1.52	$	&	$	1.53	$	&	$	39.33	$	&	$	-	$	&	$	-	$	&	$	-	$	&	$	-	$	&	$	-	$	&	$	-	$	\\
WASP-13b	&	$	0.46	$	&	$	1.21	$	&	$	4.35	$	&	$	0.053	$	&	$	156	$	&	G1V	&	$	-	$	&	$	-	$	&	$	4.9	$	&	$	-	$	&	$	-	$	&	$	-	$	&	$	-	$	&	$	-	$	&	$	-	$	\\
WASP-31b	&	$	0.50	$	&	$	1.60	$	&	$	3.50	$	&	$	-	$	&	$	-	$	&	$-$	&	$	-	$	&	$	-	$	&	$	-	$	&	$	-	$	&	$	-	$	&	$	-	$	&	$	-	$	&	$	-	$	&	$	-	$	\\
\hline
\end{tabular}
\end{center}
\end{table*}

\end{document}